# pytest-inline: An Inline Testing Tool for Python

Yu Liu[1], Zachary Thurston[2], Alan Han[2], Pengyu Nie[1], Milos Gligoric[1], Owolabi Legunsen[2]
yuki.liu@utexas.edu, {zwt3, ayh9}@cornell.edu, {pynie, gligoric}@utexas.edu, legunsen@cornell.edu
[1] UT Austin, USA    [2] Cornell University, USA

*Abstract*—We present pytest-inline, the first inline testing framework for Python. We recently proposed inline tests to make it easier to test individual program statements. But, there is no framework-level support for developers to write inline tests in Python. To fill this gap, we design and implement pytest-inline as a plugin for pytest, the most popular Python testing framework. Using pytest-inline, a developer can write an inline test by assigning test inputs to variables in a target statement and specifying the expected test output. Then, pytest-inline runs each inline test and fails if the target statement's output does not match the expected output. In this paper, we describe our design of pytest-inline, the testing features that it provides, and the intended use cases. Our evaluation on inline tests that we wrote for 80 target statements from 31 open-source Python projects shows that using pytest-inline incurs negligible overhead, at 0.012x. pytest-inline is integrated into the pytest-dev organization, and a video demo is at https://www.youtube.com/watch?v=pZgiAxR_uJg.

*Index Terms*—inline tests, software testing, Python, pytest

## I. Introduction

Software testing is the main way of checking code quality, but there is a gap in today's testing frameworks: they do not support testing *individual statements*. That is, the unit tests [5], integration tests [20], and system tests [28] supported by current frameworks can be too coarse-grained or ill-suited for developer testing needs that exist at the statement level. Yet, developers may want to test statements because:

1) Single-statement bugs occur frequently [12], but unit tests often do not catch single-statement bugs [14].
2) Some statements are hard to understand or error prone, e.g., regular expressions (regexes) [19], bit manipulation [1], string manipulation [6], or collection handling [8].
3) Statements can contain complex logic, e.g., Python one-liners [18] or Java streams [10].
4) The statement that developers want to check, i.e., the *target statement*, may be buried deeply in complicated logic that is hard to check with unit tests.

Without framework-level support for testing statements, developers use *ad hoc* approaches, like (1) "printf debugging"—printing values of variables to the console to gain visibility [21], or (2) using websites or in-IDE pop-ups to test regexes [27]. These approaches are not ideal: developers wastefully add and then remove print statements, and lose mental focus and productivity to copy code to and from websites and pop-ups. Also, developers cannot easily reuse the outcomes of these approaches. Lastly, if a target statement is in privately accessible code, some developers violate core software engineering principles to enable unit testing.

We proposed *inline tests* to meet developer needs for testing statements [17]. An inline test is a statement that allows providing arbitrary inputs and test oracles for checking the immediately preceding statement that is not an inline test. Inline tests can bring the power of unit tests to the statement level, but they should not replace unit tests or debuggers [17].

We present *pytest-inline*, the first inline testing framework for Python. Using *pytest-inline*, developers can assign test inputs to variables in a target statement and use a provided API to write oracles that specify the expected outputs. Also, *pytest-inline* runs inline tests in an isolated context and does not require interpreting the whole project. To ease installation, usage, and adoption, we develop *pytest-inline* as a plugin for pytest [22], the most popular Python testing framework.

We build *pytest-inline* by extending the prototype in our original paper [17]. The original prototype supports three kinds of test oracles, setting test display names, parameterized tests, disabling tests, grouping tests by tags, and repeated tests. In *pytest-inline*, we implement more features inspired by JUnit [11] (a mature Java testing framework) that apply to inline tests: 1) five other kinds of test oracles; 2) timeout; 3) specifying test order; 4) running inline tests in parallel; and 5) specifying assumptions. *pytest-inline* has been integrated as an officially-supported pytest plugin [25].

We evaluate *pytest-inline* on 87 inline tests that we wrote for 80 target statements in 31 open-source Python projects [17]. We find that inline tests' runtime overhead is negligible, at 0.012x of unit testing time. Our user study on the original prototype showed that all nine participants find inline tests easy to write and say that most inline tests are beneficial. Our *pytest-inline* tool will enable further research on inline testing.

We make *pytest-inline* publicly available via the pytest-dev organization: https://github.com/pytest-dev/pytest-inline.

## II. Example

Fig. 1 shows an inline test for code that we simplify from google-research/bert [3]. Line 5 checks if the variable, `name`, matches a regex for a pattern that ends in a colon and at least one digit. Directly checking the regex is not easy without

```python
1  def get_assignment_map_from_checkpoint(tvars, init_c):
2    ...
3    for var in tvars:
4      name = var.name
5      m = re.match("^(.*):\\d+$", name)
6      itest().given(name, "a:0").check_eq(m.group(1), "a")
7      if m is not None:
8        name = m.group(1)
9    ...
```

Fig. 1: Example Python code with an inline test in blue.

TABLE I: *pytest-inline*'s features. The top five are in our original prototype [17]; the bottom five are new.

| | Feature | Description | Example |
|---|---|---|---|
| Prototype | display name | Provide custom inline test name | itest(test_name="check_match_name")... |
| | parameterization | Provide multiple inputs to an inline test | itest(parameterized=True).given(name,["a:0","a:1:1"]) <br> .check_eq(m.group(1), ["a", "a:1"]) |
| | repetition | Specify number of times to run an inline test | itest(repeated=2)... |
| | tags | Tag inline tests to aid filtering | itest(tag=["regex"])... <br> $ pytest --inlinetest-group="tag-name" |
| | disabling tests | Disable an inline test | itest(disabled=True)... |
| New | timeout | Fail if inline test is still running after n seconds | itest(timeout=5)... |
| | assumptions | Execute inline test only if an assumption holds | itest().assume(platform.system() == "Linux")... |
| | inline tests order | Prioritize inline tests | $ pytest --inlinetest-order="tag-name" |
| | parallel runs | Run inline tests in parallel using *pytest-xdist* [23] | $ pytest -n auto |
| | new oracles | check_neq, check_none, check_not_none, check_same, check_not_same | itest().given(name, "a:a").check_none(m) |

statement-level testing: it is in a `for` loop and the match result is not returned from the function.

The inline test that we write for Line 5 is on Line 6. Every inline test has three parts. The "Declare" (`itest()`) part tells *pytest-inline* to process the statement as an inline test. The "Assign" (`given(name,"a : 0")`) part allows providing test inputs for the variables in the target statement. Here, `"a : 0"` is the input value for `name`. Lastly, the "Assert" (`check_eq(m,"a")`) part allows specifying a test oracle. In this case, given the test input for `name`, the `m` that the target statement computes should be `"a"` for this inline test to pass.

## III. THE *pytest-inline* FRAMEWORK

### A. API

The *pytest-inline* API provides three components:

1) **Declare**. This API component, `itest()`, signals *pytest-inline* to process a statement as an inline test and allows users to optionally specify (1) a custom test name, (2) if the inline test is parameterized, (3) a number of times to run the inline test, (4) a list of tags for filtering tests, (5) if the inline test is disabled, or (6) a timeout.
2) **Assign**. This API component, `given()`, allows developers to provide test inputs for inline tests; it takes two arguments: a variable that is used in the target statement and the value that should be assigned to that variable.
3) **Assert**. This API component, `check_*`, allows developers to specify inline test oracles. The `check_eq`, `check_neq`, `check_same`, and `check_not_same` functions take the expected value and the actual value. The `check_true`, `check_false`, `check_none`, and `check_not_none` functions take only the actual value.

### B. Features

Table I lists the features that we implement in *pytest-inline* with examples. The top five rows show our original prototype's features [17], and the bottom five rows show *pytest-inline*'s new features. To build upon our original prototype, we analyze features that JUnit [11] provides and extend the *pytest-inline* API to support those that apply to inline tests.

Parameterized inline tests allow testing the same target statement on multiple pairs of inputs and outputs. Timeout can be provided so that *pytest-inline* terminates after a specified duration (which can be useful, e.g., if there is an infinite loop). Running inline tests in parallel can save time. Developers can specify names, tags, and test orders to organize their inline tests. Tags can be used for marking and filtering tests. An inline test can have multiple tags but only one display name. We also add new features to allow developers to specify how many times to repeat an inline test and whether to temporarily disable inline tests. Repeating inline tests is useful for detecting flaky tests [2], [13]. Disabling tests can be used to skip failing tests until the fault is fixed. With assumptions, inline tests only run if a pre-condition holds. Finally, we add five new kinds of test oracles for convenience.

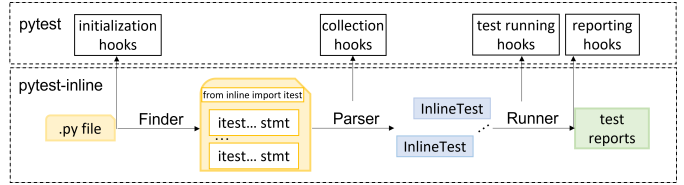

Fig. 2: Architecture of *pytest-inline*.

### C. Implementation

Fig. 2 shows the architecture of *pytest-inline*; it has three components: (1) FINDER, (2) PARSER, and (3) RUNNER.

**FINDER**. It obtains the abstract syntax tree (AST) from a given source file (using Python's AST library [26]) and locates imports of `itest` and statements that start with `itest()`.

**PARSER**. Given the output of FINDER, PARSER first traverses the AST to discover each inline test and its target statement—the first non-inline-test statement that precedes the inline test. Then PARSER (1) extracts values assigned to the arguments in the `itest()` constructor, (2) extracts the assumption in `assume()` if it exists, (3) constructs an assignment statement from each `given()`, (4) constructs an assertion statement from each `check_*()`. PARSER throws a `MalformedException` if *pytest-inline*'s API is misused. Lastly, PARSER constructs a program encapsulating the inline test with the parsed assignment statements, target statement, and assertion statements. If there is an assumption, PARSER wraps the program in an if statement with the assumption as the condition.

**RUNNER**. Given the program encapsulating each inline test from PARSER, RUNNER executes the program in an isolated

```
(inline-dev) liuyu@luzhou:~/bert$ pytest modeling.py
=========================== test session starts ============================
platform linux -- Python 3.9.15, pytest-7.2.0, pluggy-1.0.0
rootdir: /home/liuyu/bert
plugins: inline-0.1.0
collected 1 item

modeling.py .                                                        [100%]

============================ 1 passed in 0.02s =============================
```

Fig. 3: Sample *pytest-inline* output when an inline test passes.

```
(inline-dev) liuyu@luzhou:~/bert$ pytest modeling.py
=========================== test session starts ============================
platform linux -- Python 3.9.15, pytest-7.2.0, pluggy-1.0.0
rootdir: /home/liuyu/bert
plugins: inline-0.1.0
collected 1 item

modeling.py F                                                        [100%]

================================= FAILURES =================================
____________________________ [inlinetest] line326 __________________________
<ast>:7: AssertionError
E   AssertionError: m.group(1) == 'aa'
E   Actual: a
E   Expected: aa
========================= short test summary info ==========================
FAILED modeling.py::line326 - AssertionError: m.group(1) == 'aa'
============================ 1 failed in 0.06s =============================
```

Fig. 4: Sample *pytest-inline* output when an inline test fails.

context containing only the local variables that it needs and produces a pass/fail test outcome. RUNNER automatically imports libraries required by the program (e.g., re), so developers need not write import statements in inline tests. The test outcome is formatted as standard pytest output, e.g., Fig. 3.

**Integration with pytest**. We use pytest's hook functions to implement *pytest-inline* as a pytest plugin, namely by extending and customizing pytest's configuration, collection, running, and reporting phases. For example, we hook into the `pytest_exception_interact` function to customize error reporting to pretty-print the failing assertion, expected output, and actual output instead of a long stack trace. Fig. 4 shows the output of a failing inline test, obtained by changing `check_eq(m.group(1),"a")` to `check_eq(m.group(1),"aa")` in Fig. 1. Fig. 2 shows other hooks that *pytest-inline* uses.

## IV. INSTALLATION AND USAGE

**Installation**. We recommend Conda [4] for installing pytest and *pytest-inline*. A Conda environment with Python 3.9 can be created like so (pytest requires Python 3.7 or higher):

```
$ conda create --name inlinetest python=3.9 pip -y
$ conda activate inlinetest
```

Next, install pytest and *pytest-inline* in the Conda environment:

```
$ pip install pytest-inline
```

**Usage**. By default, pytest recursively discovers and runs all "test_*.py" or "*_test.py" files in the current directory. *pytest-inline* also recursively processes all ".py" files in the current directory. Users can specify what files to process, e.g., to run inline tests in ".py" files that start with "a":

```
$ pytest a*.py
```

Use `inlinetest-group` to run tagged inline tests:

```
# run only the tests with tags "str" and "bit"
$ pytest --inlinetest-group="str" --inlinetest-group="bit"
```

The `-k` option allows specifying inline tests to run by name:

```
$ pytest -k "add" # run the inline tests whose names match
    the given string expression
```

Inline tests can be run in three modes: default, inlinetest-only, and inlinetest-disable. The default mode runs inline tests and unit tests; inlinetest-only mode runs only inline tests; and inlinetest-disable mode skips inline tests but runs unit tests:

```
$ pytest # run all tests
$ pytest --inlinetest-only    # run only inline tests
$ pytest --inlinetest-disable # skip inline tests
```

When collecting inline tests, *pytest-inline* imports dependencies and throws an error if those dependencies are not installed. Users can use `inlinetest-ignore-import-errors` to ignore such errors and skip the collection of the affected files (doing so also skips the inline tests in those files):

```
$ pytest --inlinetest-ignore-import-errors
```

The default line-number order of running inline tests can be overridden using tags and `inlinetest-order`:

```
# run test tagged "str", then "bit", and then the rest
$ pytest --inlinetest-order="str" --inlinetest-order="bit"
```

Inline tests can be run in parallel after installing *pytest-xdist* [23] by using `-n` to specify the number of processes.

```
$ pip install pytest-xdist
$ pytest -n 4 # run tests in parallel with 4 processes
$ pytest -n auto # run tests in parallel with all CPU cores
```

Lastly, to generate HTML test reports, users can use the pytest-html plugin and the `html` option:

```
$ pip install pytest-html
$ pytest --html=report.html
```

## V. EVALUATION

We evaluate *pytest-inline*'s performance using the same environment to run experiments as in our original paper [17].

**Standalone experiments**. To measure the cost of running inline tests, we run the same inline tests as in our original paper [17]. These are 87 inline tests that we manually wrote for 80 target statements in 31 Python projects. Since inline tests are still new and not abundant on open-source projects, it is hard to assess *pytest-inline* costs as the number of inline tests grows. For now, we simulate such costs as in our original paper: by duplicating each inline test 10, 100, and 1000 times.

Table II shows the times to run *pytest-inline* with varying number of tests. Without duplication, the average time per inline test is 0.094s. With duplication, the average time per inline test gradually reduces to 0.001s, likely for two reasons. First, the cost of discovering inline tests is amortized with duplication, so the actual cost per inline test could be slightly higher. Second, repeatedly running an inline test benefits from reduced warm-up time. The total time to run all inline tests is almost constant when we duplicate each inline test 10 or 100 times, but that time grows greatly when we duplicate 1000 times. This dramatic growth suggests that regression testing techniques [7], [9], [15], [16], [29] will be needed to reduce inline testing costs. *pytest-inline* can be the basis on which to build those regression testing techniques. Overall, we conclude that the overhead of running these inline tests is tiny.

**Integrated experiments**. We also measure the overhead of running inline tests and unit tests together in the runtime environment specified by each project. To do so, we run inline tests and unit tests four times. The first run is for warm-up, and we average the times for the last three runs. Among 31 Python projects in our original paper, we choose the ten whose unit testing environment we can successfully configure with Python 3.7 or greater (as required by pytest): bokeh/bokeh, RaRe-Technologies/gensim, geekcomputers/Python, joke2k/faker, mitmproxy/mitmproxy, numpy/numpy, pandas-dev/pandas, psf/black, pypa/pipenv, and scrapy/scrapy. Table III shows the results. There, $O_{ITE}$ is the overhead when inline tests are enabled and run with unit tests. Without duplication, the overhead per inline test is negligible, at 0.012x. The overhead is similar with duplication. For example, when duplicating inline tests 1000 times, which brings the number of inline tests close to that of unit tests, the overhead is 0.062x.

TABLE II: Results of standalone experiments. **Dup** = duplication count, **#IT**= total no. of inline tests, $T_{IT}$[s]= total inline tests run time, $t_{IT}$[s]= average run time per inline test.

| Dup   | #IT    | $T_{IT}$[s] | $t_{IT}$[s] |
|-------|--------|-------------|-------------|
| x1    | 87     | 8.21        | 0.094       |
| x10   | 870    | 8.84        | 0.010       |
| x100  | 8,700  | 15.21       | 0.002       |
| x1000 | 87,000 | 120.17      | 0.001       |

TABLE III: Results of integrated experiments. **Dup** = duplication times, **#UT**= total no. of unit tests, **#IT**= total no. of inline tests, $T_{UT}$ [s]= total time to run unit tests, $T_{ITE}$ [s]= total time to run unit tests with inline tests enabled, $O_{ITE}$= overhead of running unit tests with inline tests enabled, $T_{ITD}$[s]= total time to run unit tests with inline tests disabled, $O_{ITD}$= overhead of running unit tests with inline tests disabled.

| Dup   | #UT     | #IT    | $T_{UT}$ [s] | $T_{ITE}$ [s] | $O_{ITE}$ | $T_{ITD}$[s] | $O_{ITD}$ |
|-------|---------|--------|--------------|---------------|-----------|--------------|-----------|
| x1    | 160,111 | 27     | 599.09       | 606.19        | 0.012     | 601.22       | 0.004     |
| x10   | 160,111 | 270    | 603.29       | 607.20        | 0.006     | 601.94       | -0.002    |
| x100  | 160,112 | 2,700  | 593.93       | 638.02        | 0.074     | 630.42       | 0.061     |
| x1000 | 160,113 | 27,000 | 649.53       | 689.50        | 0.062     | 640.93       | -0.013    |

**On user perceptions**. The user study that we performed using the original Python prototype [17] showed that participants found inline testing easy to use and beneficial. Now that we released *pytest-inline*, and have it integrated as an official pytest plugin for developers and researchers to use, we will be able to continuously obtain user feedback. For example, based on the feedback from pytest developers, we renamed the constructor from `Here` to `itest`, which is more pythonic.

## VI. CONCLUSION AND FUTURE WORK

We presented *pytest-inline* for writing inline tests in Python. We implemented *pytest-inline* as a pytest plugin, and it has been integrated into pytest-dev [24] as an official and community-maintained plugin. Our performance evaluation of *pytest-inline* showed that the cost of running inline tests is negligible, and our original prototype helped find two accepted bugs. In the future, we will add more features to *pytest-inline* based on user feedback, and use it to advance research on inline testing. *pytest-inline* could also be integrated with other pytest plugins such as pytest-mock to perform inline testing of statements that require data from files, databases, etc.

ACKNOWLEDGMENTS

We thank Nader Al Awar, Darko Marinov, August Shi, Aditya Thimmaiah, Zhiqiang Zang, Jiyang Zhang and the anonymous reviewers for their feedback on this work. This work was partially supported by a Google Faculty Research Award and the US National Science Foundation under Grant Nos. 1652517, 2019277, 2045596, 2107291, 2217696.

REFERENCES

[1] S. Bae, "Bit manipulation," in *JavaScript Data Structures and Algorithms*, 2019, pp. 339–349.
[2] J. Bell, O. Legunsen, M. Hilton, L. Eloussi, T. Yung, and D. Marinov, "DeFlaker: Automatically detecting flaky tests," in *ICSE*, 2018, pp. 433–444.
[3] "Bert," https://github.com/google-research/bert.
[4] "Conda," https://docs.conda.io/projects/conda/en/stable.
[5] E. Daka and G. Fraser, "A survey on unit testing practices and problems," in *ISSRE*, 2014, pp. 201–211.
[6] A. Eghbali and M. Pradel, "No strings attached: An empirical study of string-related software bugs," in *ASE*, 2020, pp. 956–967.
[7] M. Gligoric, L. Eloussi, and D. Marinov, "Practical regression test selection with dynamic file dependencies," in *ISSTA*, 2015, pp. 211–222.
[8] M. Gruber, S. Lukasczyk, F. Kroiß, and G. Fraser, "An empirical study of flaky tests in Python," in *ICST*, 2021, pp. 148–158.
[9] A. Gyori, O. Legunsen, F. Hariri, and D. Marinov, "Evaluating regression test selection opportunities in a very large open-source ecosystem," in *ISSRE*, 2018, pp. 112–122.
[10] "Java stream api," https://docs.oracle.com/javase/8/docs/api/java/util/stream/Stream.html.
[11] "Junit5," https://junit.org/junit5/.
[12] A. V. Kamienski, L. Palechor, C.-P. Bezemer, and A. Hindle, "PySStuBs: Characterizing single-statement bugs in popular open-source python projects," in *MSR*, 2021, pp. 520–524.
[13] W. Lam, R. Oei, A. Shi, D. Marinov, and T. Xie, "iDFlakies: A framework for detecting and partially classifying flaky tests," in *ICST*, 2019, pp. 312–322.
[14] J. Latendresse, R. Abdalkareem, D. E. Costa, and E. Shihab, "How effective is continuous integration in indicating single-statement bugs?" in *MSR*, 2021, pp. 500–504.
[15] O. Legunsen, F. Hariri, A. Shi, Y. Lu, L. Zhang, and D. Marinov, "An extensive study of static regression test selection in modern software evolution," in *FSE*, 2016, pp. 583–594.
[16] O. Legunsen, A. Shi, and D. Marinov, "STARTS: STAtic regression test selection," in *ASE Demo*, 2017, pp. 949–954.
[17] Y. Liu, P. Nie, O. Legunsen, and M. Gligoric, "Inline tests," in *ASE*, 2022, pp. 1–13.
[18] C. Mayer, *Python One-Liners: Write Concise, Eloquent Python Like a Professional*. No Starch Press, 2020.
[19] L. G. Michael, J. Donohue, J. C. Davis, D. Lee, and F. Servant, "Regexes are hard: Decision-making, difficulties, and risks in programming regular expressions," in *ASE*, 2019, pp. 415–426.
[20] A. Orso, "Integration testing of object-oriented software," p. 119, 1998.
[21] M. Perscheid, B. Siegmund, M. Taeumel, and R. Hirschfeld, "Studying the advancement in debugging practice of professional software developers," *SQJ*, vol. 25, no. 1, pp. 83–110, 2017.
[22] "Pytest," https://docs.pytest.org/en/7.2.x.
[23] "Pytest-xdist," https://github.com/pytest-dev/pytest-xdist.
[24] "pytest-dev," https://github.com/pytest-dev.
[25] "pytest-inline," https://pypi.org/project/pytest-inline.
[26] "Python ast library," https://github.com/python/cpython/blob/main/Lib/ast.py.
[27] "RegEx101," https://regex101.com.
[28] W. Tsai, X. Bai, R. Paul, W. Shao, and V. Agarwal, "End-to-end integration testing design," in *COMPSAC*, 2001, pp. 166–171.
[29] J. Zhang, Y. Liu, M. Gligoric, O. Legunsen, and A. Shi, "Comparing and combining analysis-based and learning-based regression test selection," in *AST*, 2022, pp. 17–28.